\documentclass[aps,preprint]{revtex4}%
\usepackage{amsfonts}
\usepackage{amsmath}
\usepackage{amssymb}
\usepackage{graphicx}%
\providecommand{\U}[1]{\protect\rule{.1in}{.1in}}

\begin{document}
\preprint{HEP/123-qed}
\title{On electron channeling and the de Broglie internal clock}
\author{M. Bauer}
\affiliation{Instituto de F\'{\i}sica, Universidad Nacional Aut\'{o}noma de M\'{e}xico}
\affiliation{A.P. 20-364, 01000 M\'{e}xico, D.F., MEXICO }
\pacs{PACS number}

\begin{abstract}
Electron channeling in silicon crystals has brought forward the possibility of
having detected the particle's "de Broglie internal clock", as giving rise to
the observed resonance peak at the center of the expected transmission
probability dip. A classical multiple scattering calculation fails to
represent the experimental results unless, surprisingly, the interaction
frequency is twice the de Broglie's clock frequency, that is, the
"Zitterbewegung" frequency. In the present paper, the observed characteristics
of this process are shown to be consistent with a free particle quantum
mechanical motion described by Dirac's Hamiltonian.

\end{abstract}
\date{}
\maketitle

\section{Introduction}

Electron channeling in silicon crystals\cite{Catillon,Gouanere} has brought
forward the possibility of having detected a particle's \textquotedblleft
internal clock\textquotedblright, as an intrinsic oscillation whose frequency
is given by de Broglie's daring association $h\nu=m_{0}c^{2}$, where $h$ is
Planck's constant, $m_{0}$ is the particle rest mass and $c$ the speed of
light in vacuum.\cite{Broglie} More recently, a clock linked to this relation
- referred as "de Broglie internal clock", "Compton clock" or "de Broglie
periodic phenomenon" - has been demonstrated using an optical frequency to
self-reference a Ramsey-Bord\'{e} atom interferometer.\cite{Lan}

The cited channeling experiments in which relativistic electrons are aligned
along a major axial direction of a thin single crystal, do exhibit a reduced
transmission probability with respect to neighboring angles. At certain
energies however, a central sharp peak appears in the atomic row direction.
The pattern observed is a \textquotedblleft W\textquotedblright\ instead of a
\textquotedblleft U\textquotedblright. This central peak is attributed to a
resonance process denoted as \textquotedblleft rosette
motion\textquotedblright\ , which results in a reduction of the multiple
scattering effects for electrons moving parallel to a string of atoms with a
momentum such that they pass atoms with a frequency equal to the de Broglie
frequency. The expected consequence is a higher transmittivity relative to
closely nearby directions and momenta, as a fraction of the electrons are
trapped in a spiral motion about the atomic row which, projected onto the
transverse plane, executes rosettes, i.e., bound orbits that precess.

A phenomenological calculation by a Montecarlo method is carried
out\cite{Catillon,Gouanere}, in which the electron motion is described
classically, based on a theoretical model applied successfully to the
channeling process at relativistic energies\cite{Lindhard,Gemmell}. The basic
assumption made is that the distance $L$ travelled during a de Broglie
laboratory period is equal to the interatomic distance $d$ in the crystal row.
This defines the goup velocity $v_{gp}$ and consequently the energy. Although
giving an electron energy close to the experimental one, the theoretical
calculation does not yield the experimental results. It is then noted that,
surprisingly, the results are reproduced at the same energy\ if $L$ is taken
to be equal to twice the distance $d$. It is noted that this requires to
consider a reduced period and consequently a higher frequency, namely the
\textquotedblleft Zitterbewegung\textquotedblright\ frequency.

In the present paper it is shown that all the observed characteristics of this
process follow naturally from the quantum mechanical free-particle motion
described by a Dirac Hamiltonian (albeit with an effective mass resulting from
the average interaction with the crystal atoms). A posible explanation of how
the surprising fit arises is presented here.

\section{The free particle Dirac Hamiltonian as a symmetry operation}

Consider the free particle Dirac Hamiltonian
\begin{equation}
H=c\mathbf{\boldsymbol{\alpha}.p}+\beta m_{0}c^{2}%
\end{equation}
where \textbf{$\alpha$}$=(\mathbf{\alpha}_{x},\mathbf{\alpha}_{y}%
,\mathbf{\alpha}_{z})$ and $\beta$ are the Dirac matrices. Recalling that the
infinitesimal $(\epsilon\ll1)$ unitary operator $S(\epsilon)=e^{\{i\epsilon
\mathbf{p}/\hbar\}}$ acting on a position eigenstate yields a displaced
eigenstate, namely\cite{Messiah}:%
\[
S(\epsilon)|x>=e^{\{i\epsilon\mathbf{p}/\hbar\}}|x>=[1+(i\epsilon
\mathbf{p}/\hbar)+%
\frac12
(i\epsilon\mathbf{p}/\hbar)%
{{}^2}%
+.....]|x>=|x+\epsilon>,
\]
it follows that the infinitesimal unitary operator $(\tau\ll1)$:%
\begin{align}
U(\tau)  &  =e^{-i\tau H/\hbar}=e^{-i\tau\{c\mathbf{\boldsymbol{\alpha}%
.p}+\mathbf{\beta}m_{0}c^{2}\}/\hbar}=\nonumber\\
&  =[1+(i\tau c\mathbf{\boldsymbol{\alpha}.p}/\hbar)+%
\frac12
(i\tau c\mathbf{\boldsymbol{\alpha}.p}/\hbar)%
{{}^2}%
+.....]exp^{-i\beta\tau m_{0}c^{2}/\hbar}%
\end{align}
induces, in configuration space, a position displacement by an amount
$\delta\mathbf{r}=\tau c$\textbf{$\boldsymbol{\alpha}$} and a phase shift
$\delta\phi=\beta(\tau m_{0}c^{2}/\hbar)$, i.e.:%
\begin{equation}
\Phi(\mathbf{r})=<\mathbf{r}\mid\Phi>=e^{i\phi}\varphi(\mathbf{r})\rightarrow
e^{i(\phi+\mathbf{\delta}\phi)}\varphi(\mathbf{r}+\tau
c\mathbf{\boldsymbol{\alpha}})
\end{equation}
\qquad\ As $[U(\tau),H]=0$, the displaced wave function satisfies the same
Schr\"{o}dinger equation. $U(\tau)$ is thus a symmetry operation. Finite
displacements are achieved by repeated applications. The phase shift is seen
to be related to the reduced de Broglie (or Compton) frequency $(m_{0}%
c^{2}/\hbar)$.

The Dirac relativistic free particle motion in the Heisenberg picture is given
by:%
\begin{equation}
\mathbf{r}(t)=\mathbf{r}(0)+(c^{2}\mathbf{p}/H)t+(\hbar c/2iH)(e^{2iHt/\hbar
}-1)\mathbf{F}%
\end{equation}
with $\mathbf{F}=c\mathbf{\alpha}-(c^{2}\mathbf{p}/H)$ \ Averaging over a
general positive energy wave packet one has $<c^{2}\mathbf{p}/H>=$%
\ $\mathbf{v}_{gp}$, the group velocity, and:%
\begin{equation}
<\mathbf{r}(t)>=<\mathbf{r}(0)>+\mathbf{v}_{gp}t+<(\hbar c/2iH)(e^{2iHt/\hbar
}-1)\mathbf{F>}%
\end{equation}
Thus, as pointed out by Schr\"{o}dinger, $<\mathbf{r}(t)>$ is an oscillatory
motion ("Zitterbewegung") about the rectilinear uniform motion $<\mathbf{r}%
(0)>+\;\mathbf{v}_{gp}t$. The oscillation arises when both positive and
negative energies are present in the wave packet, bringing in a contribution
of the operator $\mathbf{F}$\cite{Thaller,Greiner}.\ The position coincides
with the uniform motion at times $t_{n}$ such that $2Ht_{n}$/$\hslash=2\pi n$
\ with $\ n=1,2,...$, or equivalently at intervals $\Delta t$ $=t_{n+1}%
-t_{n}=2\pi\hslash/2H=h/2H=h/2m_{0}c^{2}\gamma$, where $\gamma=[1-(\mathbf{v}%
_{gp}/c)^{2}]^{-1/2}$ is the Lorentz factor. This interval corresponds to the
Zitterbewegung laboratory period, (denoted as Z period) $T_{Z}^{lab}%
=(1/\nu_{Z}^{lab})$, thus equal to half the laboratory de Broglie period
(denoted as B period).

The phase at time $t$ is:%
\begin{equation}
<\phi(t)>=<\beta(t)>t\text{ }(m_{0}c^{2}/\hbar)
\end{equation}
where $<\beta(t)>=<m_{0}c^{2}/H+e^{2iHt/\hbar}G>$ with $G=\beta(0)-(m_{0}%
c^{2}/H)$ \ \ , an operator that also connects only positive and negative
energy eigenstates. Then at the coincidence times $t_{n}=2\pi n\hbar/2H$, one
has $<\beta(t_{n})>=<m_{0}c^{2}/H+G>=<\beta(0)>$ and the phase is:%
\begin{equation}
<\phi(t_{n})>=<\beta(t_{n})>t_{n}\text{ }(m_{0}c^{2}/\hbar)=<\beta(0)>(2\pi
n\hbar/2H)(m_{0}c^{2}/\hbar)=<\beta(0)>(n\pi/\gamma)
\end{equation}
Consequently the change in phase bettween two consecutive coincidence times is
given by:%
\begin{equation}
\Delta\varphi=<\beta(0)>(\pi/\gamma)\approxeq0\ \ \
\end{equation}
when$\ v_{gp}\;$approaches $c.$Thus, at each coincidence time, a relativistic
electron arrives with the same amplitude and phase.

\section{Electron Channeling}

In a channeling set-up the rectilinear motion would be aligned along the
crystal atomic row. Then the electron may be made to coincide with crystal
atoms given the appropiate group velocity and consequently the appropiate
energy. In the experiment cited\cite{Catillon,Gouanere}, the requirement was
made that the electron should advance the interatomic distance $\ d$ \ in one
B period, i.e., $L=v_{gp}T_{B}^{lab}=d$ . Now $T_{B}^{lab}=$ $(h\gamma
/m_{0}c^{2})$ \ since the de Broglie linear frequency in the laboratory is
$\nu_{B}^{lab}=$ $(m_{0}c^{2}/h\gamma)$.\cite{Broglie,Lochak}. Then:%

\begin{equation}
v_{gp}=(d/T_{B}^{lab})=(d)(m_{0}c^{2}/h\gamma)
\end{equation}
Defining $d(m_{0}c^{2}/hc)=\alpha$ and \ $(v_{gp}/c)=\beta$ , it follows that:%

\[
(\beta\gamma)^{2}=\beta^{2}(1-\beta)^{-1}=\alpha^{2},
\]
and finally%
\begin{equation}
\beta=(v_{gp}/c)=\alpha/(\alpha+1)^{1/2}\approx1-1/(2\alpha^{2})\text{
\ }if\text{ }\alpha>>1
\end{equation}
In the reported experiment $\ d=3.84$ $\mathring{A}=3.84$ $(10^{5})$ $F$ .
Then%
\begin{equation}
\alpha=d(m_{0}c^{2})/hc=(1/2\pi)d(m_{0}c^{2})/\hbar c=158.2707
\end{equation}
and the motion is seen to be highly relativistic. The laboratory energy is:%
\begin{equation}
E=\mathbf{\ }m_{0}c^{2}\gamma\approx m_{0}c^{2}\alpha=m_{0}c^{2}d(m_{0}%
c^{2}/hc)=0.511(158.2707)\text{ }MeV=80.876\text{ }MeV
\end{equation}
The corresponding phase shift is from Eq.(8) close to zero, as $T_{B}%
^{lab}=2T_{Z}^{lab}$. The electron located at one atom at a certain moment
will reach the next one after one de Broglie period (or two Zitterbewegung
periods) with the same amplitude and phase. This can be the base for a
resonance phenomenon.

Note that the same energy value of $80.876$ $MeV$ results from requiring the
interatomic distance $\ d$ \ to be attained after two Z periods. On the other
hand, from the above derivation, if the full interatomic distance $d$ is to be
attained after one Z period, or equivalently $2d$ in a B period, the energy
must be $E=$ $161.752$ $MeV$. Finally, a resonance energy \ $E=40.438\;MeV$
follows from requiring the distance $d/2$ to be travelled in a B period. All
these situations are recognized in Refs.(1, 2), suggesting that the phenomenon
should also be observable at these energies.

\section{The semi-classical calculation}

As seen in Fig. 4 of Ref.(1) the classical phenomenological calculation does
not reproduce the experimental result. On the other hand, taking $L=2d$ \ ,
surprisingly seems to succeed at the same energy, whereas one would expect a
group velocity twice as large and an energy of $161.752$ \ $MeV$, as noted
above and also in Ref.1.

A possible explanation may be the following. The calculation has as
independent parametrs the intensity $K$ of confining potential $U(x_{i}%
,y_{i},z_{i})=K\;x_{i}^{2}y_{i}^{2}z_{i}^{2}$,\ and the range $\sigma$ of the
Gaussian shape used to introduce a fluctuating length at each step. The
deviation angle at each step in the $x$-direction of the transverse plane for
$L=d$ is given as :%
\begin{equation}
d\alpha_{i}=2LKy_{i}^{2}z_{i}^{2}x_{i}/m=2dKy_{i}^{2}z_{i}^{2}x_{i}/m
\end{equation}
and similarly in \ the $y$-direction\ (see Eqs.(2) and (3) of Ref. (1)). If
one considers now $L=2d$ \ one can rewrite this equation as%
\begin{equation}
d\alpha_{i}=2dK^{\prime}y_{i}^{2}z_{i}^{2}x_{i}/m
\end{equation}
with $K^{\prime}=2K$. This is equivalent to conserve $L=d$, and consequently
the group velocity and energy, but to double the strength of the potential and
thus the restoring force. This results in a contraction of all dimensions. As
such, it can be expected to reduce the width of the transmission dip.

In addition the normalized Gaussian was given the value $\sigma=L/4=d/4$. If
now $L=2d$, the range parameter becomes $\sigma^{\prime}=d/2$, increasing its
range and reducing the maximum value by $1/2$. As pointed out in Ref.(1), "the
parameter $\sigma$\ \ acts only on the dip amplitude, not on its width". Using
a more extended Gaussian function will reduce the amplitude of the
transmission dip. Thus the above changes in $K$ and $\sigma$ are expected to
reduce both the width and the amplitude of the transmission dip, bringing the
calculation in closer agreement with the experiment, as seems to have occurred
in Ref.1.

Finally, the remainig difference between the theoretical and the experimental
resonant energies could be accounted by considering that the electron's motion
in the crystal is well represented by the free-particle Dirac Hamiltonian,
albeit with an effective mass $m^{\ast}$ instead of $m_{0}$ that takes into
account the average interaction of the electron with the crystal
atoms\cite{Kittel,Green}. Following this assumption through the above
derivation yields ($m^{\ast}$/$m_{0}$)$^{2}$ $=(81.1)/(80.876)=1.00225$ and
$m^{\ast}$ $=1.0013\ m_{0}$.

\section{Conclusion}

The resonant conditions observed in the significative electron channeling
experiment considered have been shown to be consistent with Dirac's quantum
mechanical description of the free-particle motion. An explanation is offered
on how the modified calculation presented in the experimental
paper\cite{Catillon,Gouanere} came to be more in agreement with the observed
result. Notwithstanding its acceptance, it is to be stressed that the
experiment provides an indirect evidence of the Zitterbewegung phenomenon, and
of the de Broglie clock, in the case of electrons, where direct observation is
still beyond present technical capabilities. At present Zitterbewegung has
indeed been observed in experimental conditions whose dynamics simulate
Dirac's Hamiltonian\cite{Gerritsma,LeBlanc}.

\end{document}